\documentclass[conference]{IEEEtran}
\IEEEoverridecommandlockouts
\usepackage{cite}
\usepackage{amsmath,amssymb,amsfonts}
\usepackage{graphicx}
\usepackage{textcomp}
\usepackage{xcolor}
\usepackage{algorithm}
\usepackage{algpseudocode}
\usepackage{multirow}
\usepackage{tabularx}
\def\BibTeX{{\rm B\kern-.05em{\sc i\kern-.025em b}\kern-.08em
    T\kern-.1667em\lower.7ex\hbox{E}\kern-.125emX}}
\begin{document}

\title{Particle Flow Gaussian Particle Filter \\
}

\author{\IEEEauthorblockN{1\textsuperscript{st} Given Name Surname}
\IEEEauthorblockA{\textit{dept. name of organization (of Aff.)} \\
\textit{name of organization (of Aff.)}\\
City, Country \\
email address or ORCID}
\and
\IEEEauthorblockN{2\textsuperscript{nd} Given Name Surname}
\IEEEauthorblockA{\textit{dept. name of organization (of Aff.)} \\
\textit{name of organization (of Aff.)}\\
City, Country \\
email address or ORCID}
\and
\IEEEauthorblockN{3\textsuperscript{rd} Given Name Surname}
\IEEEauthorblockA{\textit{dept. name of organization (of Aff.)} \\
\textit{name of organization (of Aff.)}\\
City, Country \\
email address or ORCID}
}
\author{\IEEEauthorblockN{Karthik Comandur}
\IEEEauthorblockA{\textit{SPCRC} \\
\textit{IIIT Hyderabad}\\
Hyderabad, India \\
karthik.comandur@research.iiit.ac.in}
\and

\IEEEauthorblockN{Yunpeng Li}
\IEEEauthorblockA{\textit{Deparment of Computer Science} \\
\textit{University of Surrey}\\
Guildford, UK \\
yunpeng.li@surrey.ac.uk}
\and

\IEEEauthorblockN{Santosh Nannuru}
\IEEEauthorblockA{\textit{SPCRC} \\
\textit{IIIT Hyderabad}\\
Hyderabad, India \\
santosh.nannuru@iiit.ac.in}
}

\maketitle

\begin{abstract}
State estimation in non-linear models is performed by tracking the posterior distribution recursively. A plethora of algorithms have been proposed for this task. Among them, the Gaussian particle filter uses a weighted set of particles to construct a Gaussian approximation to the posterior. In this paper, we propose to use invertible particle flow methods, derived under the Gaussian boundary conditions for a flow equation, to generate a proposal distribution close to the posterior. The resultant particle flow Gaussian particle filter (PFGPF) algorithm retains the asymptotic properties of Gaussian particle filters, with the potential for improved state estimation performance in high-dimensional spaces. We compare the performance of PFGPF with the particle flow filters and particle flow particle filters in two challenging numerical simulation examples.

\end{abstract}

\begin{IEEEkeywords}
Particle filters, particle flow filters, Gaussian particle filters, particle flow particle filters.
\end{IEEEkeywords}

\section{Introduction}

Particle-based methods have been a popular class of sequential statement estimation techniques to approximate intractable posterior distributions. Particle filters \cite{gordon1993novel}, a.k.a. sequential Monte Carlo filters, use a set of particles and their associated weights to represent and track the state posterior. In high-dimensional state spaces or when the measurements are highly informative, the particle filter suffers from particle weight degeneracy, i.e. most particles have negligible weights \cite{bickel2008sharp,bengtsson2008curse,snyder2008obstacles}. Weight degeneracy results in poor approximation of the posterior distribution and deteriorates the performance of particle filter. The sequential importance resampling (SIR) particle filter uses resampling to alleviate weight degeneracy, but it results in a loss of diversity among the particles known as sample impoverishment~\cite{arulampalam2002tutorial}.

Numerous methods have been proposed to address the weight degeneracy issue. The auxiliary particle filter \cite{pitt1999filtering} is a variant of the SIR filter, which samples particles more effectively than the vanilla bootstrap particle filter \cite{gordon1993novel} by considering information from the new measurement. The Rao-Blackwellised particle filter \cite{doucet2000s} increases the efficiency of particle filtering by analytically marginalizing out some state variables. Though these particle filters are effective in many settings, they often perform poorly in high-dimensional state spaces or require certain structures of the underlying models. An alternate direction involves performing Markov Chain Monte Carlo (MCMC) after the resampling step in particle filters \cite{berzuini1997dynamic,gilks2001following,godsill2001improvement,musso2001improving,brockwell2010sequentially,septier2015langevin}. The methods along this direction may still suffer from largely duplicated samples in high-dimensional spaces until a large number of computationally expensive MCMC iterations are performed. 

Particle flow filters \cite{daum2007nonlinear,daum2008particle,daum2009gradient,daum2010exact,daum2013particle,daum2014seven} are a class of filtering methods designed to continuously migrate particles from prior to posterior following the Fokker-Planck equation. There is no importance sampling step in particle flow filters which avoids the weight degeneracy issue. A major drawback of particle flow filters is the lack of statistical consistency resulting from various model assumptions and approximations in numerical implementations.
To overcome this issue, invertible particle flow was constructed to generate proposal distributions within particle filtering~\cite{li2017particle} or sequential MCMC~\cite{li2019} framework. These methods acquire the desired properties of particle flow in high-dimensional filtering tasks and inherit the statistical properties of the encompassing framework. Invertible particle flow procedure developed in~\cite{li2017particle} were adapted from the exact Daum and Huang (EDH) filter~\cite{daum2010exact} and the localized exact Daum and Huang (LEDH) filter~\cite{ding2012implementation}. The EDH filter was derived with the Gaussianity assumption on the boundary conditions of the ordinary differential equation (ODE) governed by the Fokker-Planck equation. Yet, the numerical implementation of the invertible particle flow defines affine mappings~\cite{li2019}, which suggests that the generated proposal distribution can be a Gaussian or a local Gaussian approximation to the posterior.

Gaussian particle filter (GPF) \cite{article} is a variant of the particle filter where the weighted particle set is used to approximate the posterior distribution as Gaussian. The mean and covariance matrix of the Gaussian, instead of particles, are propagated through time steps. The particles are sampled from this Gaussian distribution before each iteration. The predictive distribution is also approximated as Gaussian and used in the weight update expression. Hence, the resampling step is circumvented due to the Gaussian approximation of both the predictive and posterior distributions.


In this paper, we explore the benefit of incorporating the invertible particle flow procedure into the Gaussian particle filter, due to their shared Gaussianity assumptions and approximations. By incorporating invertible particle flow into the GPF, we eliminate the need to resample particles to avoid weight degeneracy and use the encompassing importance sampling mechanism to incorporate information from non-linear non-Gaussian models. The main contribution of the paper are:
(i) we incorporte an invertible particle flow into a Gaussian particle filter to construct an effective proposal density.
(ii) we derive the modified importance sampling weight due to the incorporation of particle flow into the GPF. (iii) We evaluate and compare the performance of the proposed filter in two challenging numerical simulation setups.

The structure of the paper is as follows:
in Section \ref{sec:problem_statement} we discuss the non-linear filtering problem statement. A brief review of Gaussian particle filter and particle flow is provided in Section \ref{sec:related_work}. In Section \ref{sec:pfgpf} we describe the proposed particle flow Gaussian particle filter algorithm and its numerical implementation. Simulations and results are presented in Section \ref{sec:simulations} while the summary of the results and future scope are given in Section \ref{sec:summary}.


\section{Problem statement}
\label{sec:problem_statement}
The problem of state estimation in non-linear filtering is to estimate the unobserved state $x_t$ of the system at time step $t$ . This is performed by tracking the posterior density $p(x_t|z_{1:t})$ over time step $t$, where $z_{1:t} = \{z_1,...,z_t\}$ is set of observations collected up to $t$. The state $x_t$ and the observation $z_t$ follow the dynamic model and observation model given by \eqref{eq:dynamic_model} and \eqref{eq:measurement_model} respectively:
\begin{align} 
x_0 &\sim {p}_0(x)\,\,, \\
x_t &= g_t(x_{t-1},v_t)\,\,, \; t = 1,2,\ldots \label{eq:dynamic_model} \\ 
z_t &= h_t(x_t,w_t)\,\,, \; t = 1,2,\ldots \label{eq:measurement_model} 
\end{align}
where $p_0(x)$ is the initial state distribution, $g_t$ is the state transition function, and $h_t$ is the observation model which generates the observations $z_t$. $v_t$ and $w_t$ are the process noise and observation noise respectively.
We assume that $g_t(\cdot,0)$ is bounded and $h_t(\cdot,0)$ is a $C^1$ function, i.e. $h_t(\cdot,0)$ is differentiable everywhere and its derivatives are continuous. 

\section{Related work}
\label{sec:related_work}

\subsection{Gaussian particle filter}
The Gaussian particle filter \cite{article} approximates both the predictive and posterior distributions as Gaussian. Consider $\mathcal{N}(\cdot \,; \mu_{t-1},\Sigma_{t-1})$, a Gaussian approximation of the posterior at time step $t-1$ where $\mu_{t-1}$ and $\Sigma_{t-1}$ are the mean and covariance of the Gaussian. A set of particles $\{{x}^{i}_{t-1}\}^{N_{p}}_{i=1}$ is drawn from this Gaussian in the start of time step $t$. As in particle filters, the GPF consists of prediction and update steps. 

\subsubsection{Prediction}
A predicted set of particles $\{{x}^{i}_{t|t-1}\}^{N_{p}}_{i=1}$ is generated by propagating the particle set  $\{{x}^{i}_{t-1}\}^{N_{p}}_{i=1}$
through the dynamic model~\eqref{eq:dynamic_model}.
A Gaussian distribution $\mathcal{N}(\cdot \,;\overline{\mu}_{t},\overline{\Sigma}_{t})$ is constructed from this set of particles with the mean and covariance given by
\begin{align} 
{\overline{\mu}}_{t} &= \frac{1}{N_{p}} {\sum}^{N_{p}}_{i=1} {x}^{i}_{t|t-1} \,, \\
{\overline{\Sigma}}_{t} &= \frac{1}{N_{p}} {\sum}^{N_{p}}_{i=1} ({x}^{i}_{t|t-1}-\overline{\mu}_{t}) {({x}^{i}_{t|t-1}-\overline{\mu}_{t})}^{T}.
\end{align}

\subsubsection{Update}
Samples $x^{i}_{t}$ are drawn from a proposal distribution $\pi(\cdot)$. To account for the difference between the true posterior and the proposal distribution, importance weights are computed for each particle:
\begin{align}
{w}^{i}_t &\propto \frac{p({x}^{i}_{t}|z_{0:t-1})p(z_t|{x}^{i}_{t})}{\pi({x}^{i}_{t}|z_{0:t})} \,,
\end{align}
where $p({x}_{t}|z_{0:t-1})$ is the predictive distribution, $p(z_t|{x}_{t})$ is the likelihood, and $\pi({x}_{t}|z_{0:t})$ is the importance sampling distribution. Since the predictive distribution is approximated by the Gaussian $\mathcal{N}(\cdot \,;\overline{\mu}_{t},\overline{\Sigma}_{t})$, the weights become
\begin{align}
{w}^{i}_t \propto \frac{\mathcal{N}({x}^{i}_{t};\overline{\mu}_{t},\overline{\Sigma}_{t})p(z_t|{x}^{i}_{t})}{\pi({x}^{i}_{t}|z_{0:t})}.
\end{align}
The posterior distribution is approximated by the Gaussian $\mathcal{N}(\cdot \,; {\mu}_{t},{\Sigma}_{t})$ with the mean and covariance computed empirically from the particles and normalized weights:
\begin{align} 
{{\mu}}_{t} &= {\sum}^{N_{p}}_{i=1} {w}^{i}_{t} {x}^{i}_{t} \,, \label{eq:GPF_mean_t} \\ 
{{\Sigma}}_{t} &= {\sum}^{N_{p}}_{i=1} {w}^{i}_{t}({x}^{i}_{t}-{\mu}_{t}) {({x}^{i}_{t}-{\mu}_{t})}^{T}. \label{eq:GPF_cov_t}
\end{align}
The above estimates ${{\mu}}_{t}$ and ${{\Sigma}}_{t}$ converge asymptotically (i.e., as $N_p \to \infty$) to the minimum mean square error (MMSE) estimates of posterior mean and covariance i.e., $\overline{\mu}_t = E[x_t|z_{0:t}]$ and $E[(x_t - \overline{\mu}_t)(x_t - \overline{\mu}_t)^T|z_{0:t}]$ respectively (see Theorem 1 and Corollary 1 in \cite{article}).
The proposal distribution $\pi(\cdot)$ can be any user-specified distribution, e.g. the predictive distribution. In this work we use invertible particle flow as the proposal distribution. A brief review of particle flow is provided next.

\subsection{Particle flow}
Particle flow methods migrate the particles from the prior to the posterior using a flow equation. Consider a set of $N_p$ particles $\{x_{t-1}^i\}_{i=1}^{N_p}$ which approximate the posterior distribution at time $t-1$. These particles are propagated through the dynamic model to generate the predicted set of particles representing the predicted state distribution at time step $t$.

Particle flow then migrates predicted particles to the posterior distribution. The underlying flow process is modeled as a background stochastic process $\eta_\lambda$ in a pseudo time interval $\lambda \in [0,1]$. For brevity, the time index $t$ is not included in this section. There are two main types of particle flow: deterministic flow \cite{daum2010exact,daum2010exact1} which involves no particle diffusion, and stochastic flow \cite{daum2013particle} which includes particle diffusion. The deterministic flow equation is given by an ordinary differential equation,
\begin{align}
 \frac{d{\eta}^i_\lambda}{d\lambda} &= f({\eta}^i_\lambda,\lambda)  \,\,,
\end{align}
where the function $f$ is governed by the Fokker-Planck equation and additional flow constraints \cite{daum2010exact1}. Commonly used deterministic particle flow, such as the exact Daum and Huang (EDH) filter \cite{daum2010exact} and the localized exact Daum and Huang (LEDH) filter \cite{ding2012implementation}, are reviewed below.

\subsubsection{Exact Daum and Huang filter}
The flow equation for the exact Daum and Huang filter is given by 
\begin{align}
f({\eta}^i_\lambda,\lambda) = A(\lambda){\eta}^i_\lambda + b(\lambda)\,\,,
\end{align}
where
\begin{align}
&A(\lambda) = -\frac{1}{2}P{H(\lambda)}^T{(\lambda H(\lambda)P{H(\lambda)}^T+R)}^{-1} H(\lambda) \,\,,\label{edh_eqA} \\
&b(\lambda) = (I+2\lambda A(\lambda)) \times \nonumber \\
& \; [(I+\lambda A(\lambda)) P{H(\lambda)}^T{R}^{-1}(z-e((\lambda)) + A(\lambda)\overline{\eta_{0}}] \,\,, \label{edh_eqB}
\end{align}
where $\overline{\eta}_{0}$ and $P$ are the predicted mean and covariance, respectively, and $R$ is the observation covariance matrix. For linear observation models, $H(\lambda)$ is the measurement matrix. For non-linear observation models, linearization is performed at the mean $\overline{\eta_\lambda}$ of the intermediate distribution to obtain the Jacobian matrix $H(\lambda) = \frac{\partial h(\eta,0)}{\partial \eta}
{\Big|}_{\eta=\overline{\eta}_\lambda}$ and $e(\lambda)$ is given by
\begin{align}
e(\lambda) &= h(\overline{\eta}_{\lambda},0) - H(\lambda) \overline{\eta}_{\lambda} \,.
\end{align} 
In the EDH filter, the flow parameters $A(\lambda)$ and $b(\lambda)$ are the same for all the particles.


\subsubsection{Localized exact Daum and Huang filter}
The flow parameters in the localized exact Daum and Huang (LEDH) filter are computed individually for each particle. For the $i$-th particle,
\begin{align}
&{A}^i(\lambda) = -\frac{1}{2}P{{H}^i(\lambda)}^T{(\lambda {H}^i(\lambda)P{{H}^i(\lambda)}^T+R)}^{-1} {H}^i(\lambda) \,\,, \label{ledh_eqA} \\
&{b}^i(\lambda) = (I+2\lambda {A}^i(\lambda)) \times \nonumber\\
& \; [(I+\lambda {A}^i(\lambda)) P{{H}^i(\lambda)}^T{R}^{-1}(z-{e}^i(\lambda) + {A}^i(\lambda)\overline{\eta}_0]\,\,.
\label{ledh_eqB}
\end{align}
Here the linearization of the observation model is performed at each individual particle as ${H}^i(\lambda)=\frac{\partial h(\eta,0)}{\partial \eta} {\Big|}_{\eta={{\eta}^i_\lambda}}$ and 
$e^{i}(\lambda) = h(\eta^{i}_{\lambda},0)-H^{i}(\lambda) \eta^{i}_{\lambda}$.


\section{Particle flow Gaussian particle filter}
\label{sec:pfgpf}
We now propose to incorporate particle flow into the Gaussian particle filter and present the numerical implementation details. 

\subsection{The prediction step}
Consider the particle set $\{{x}^i_{t-1}\}^{N_p}_{i=1} \sim \mathcal{N}(x_{t-1};\mu_{t-1},\Sigma_{t-1})$ which approximates the posterior at the time step $t-1$. This particle set is propagated through the dynamic model to generate a new set of particles named the predictive particles:
\begin{align}
{\eta}^i_{0} &= {g}_t({x}^i_{t-1},v_t).
\end{align}
The mean and covariance of this set of particles are used for the Gaussian $\mathcal{N}(\cdot \,;\overline{\mu}_{t|t-1},\overline{\Sigma}_{t|t-1})$ to approximate the predictive distribution.
\begin{align} 
{\overline{\mu}}_{t|t-1} &= \frac{1}{N_{p}} {\sum}^{N_{p}}_{i=1} {\eta}^{i}_{0} \,\,, \\
{\overline{\Sigma}}_{t|t-1} &= \frac{1}{N_{p}} {\sum}^{N_{p}}_{i=1} ({\eta}^{i}_{0}-{\overline{\mu}}_{t|t-1}) {({\eta}^{i}_{0}-{\overline{\mu}}_{t|t-1})}^{T}\,\,.
\end{align} 

Following the practice in~\cite{li2017particle}, auxiliary particle flow is performed for particles generated through dynamic model without noise $\{{\bar{\eta}}^i_{0}\}^{N_p}_{i=1}$, where
\begin{align}
{\bar{\eta}}^i_{0} = {g}_t({x}^i_{t-1},0)\,\,.
\end{align}
This ensures that the linearization is performed at deterministic particles to generate flow parameters through Equations~\eqref{edh_eqA} and~\eqref{edh_eqB} for the EDH flow and Equations \eqref{ledh_eqA} and~\eqref{ledh_eqB} for the LEDH flow. The stored flow parameters are then applied to the predicted particle set to generate the set $\{{\eta}^i_{1}\}^{N_p}_{i=1}$ as samples from the proposal distribution. This particle flow process leads to invertible mapping of particles under mild assumptions on the model and numerical implementation~\cite{li2017particle}.


\subsection{The update step}
With invertible particle flow, the importance sampling distribution $\pi(\cdot)$ is computed as 
\begin{align}
\pi({\eta}^{i}_{1}|z_{0:t}) &= \frac{\mathcal{N}(\eta^{i}_1;\overline{\mu}_{t|t-1},\overline{\Sigma}_{t|t-1})}{|\dot{T}{({\eta}^{i}_{0};z_{t},{x}^{i}_{t-1})}|}\,\,,
\label{eq:PFGPF_IS}
\end{align}
where $T(\cdot)$ is the mapping function between the particles before and after the flow. The denominator in~\eqref{eq:PFGPF_IS} is the absolute value of the Jacobian determinant of the transport mapping function $T(\cdot)$ given by~\cite{li2017particle}
\begin{align}
|T({\eta}^i_0:z_t,{x}^i_{t-1})| &= |\text{det}(\frac{d{\eta}^i_1)}{d{\eta}^i_0)})| \\
&= {\Pi}^{N_{\lambda}}_{j=1} |\text{det}(I+\epsilon_j{A}^i_j(\lambda))| \,\,,
\end{align}
where $N_{\lambda}$ is the number of pseudo time steps.  

The importance weights of the $i$-th particle $\{{\eta}^i_{1}\}^{N_p}_{i=1}$ is computed as
%
%
\begin{align}
{w}^{i}_t &\propto \frac{\mathcal{N}(\eta^{i}_1;\overline{\mu}_{t|t-1},\overline{\Sigma}_{t|t-1})p(z_t|{\eta}^{i}_{1})|\dot{T}{({\eta}^{i}_{0};z_{t},{x}^{i}_{t-1})}|}{\mathcal{N}(\eta^{i}_0;\overline{\mu}_{t|t-1},\overline{\Sigma}_{t|t-1})}\,\,.
\label{eq:PFGPF_weight}
\end{align}
In the GPF framework, the Gaussian distribution $\mathcal{N}(x_t;\mu_{t},\Sigma_{t})$ is used to approximate the posterior distribution after the weight update step, where the weighted mean $\mu_{t}$ and weighted covariance matrix are computed as in \eqref{eq:GPF_mean_t} and \eqref{eq:GPF_cov_t} with $x^{i}_{t}$ replaced by $\eta^{i}_{1}$. As in~\cite{li2017particle}, the extended Kalman filter (EKF) or the unscented Kalman filter (UKF) \cite{julier1997new} is used to estimate the predictive covariance matrix needed in estimating the flow parameters. Algorithm~\ref{Particle Flow Gaussian Particle Filter} outlines the proposed filter where we use the LEDH flow for migrating the particles from the prior to the posterior. Since PFGPF is essentially a specific case of GPF, it inherits the asymptotic proprieties of the GPF~\cite{article}.

\begin{algorithm}
\linespread{1.1}\selectfont
\caption{{ Particle flow Gaussian particle filter (LEDH)}}
\label{Particle Flow Gaussian Particle Filter}
\begin{algorithmic}[1]
\label{alg:PFGPF}
\State  Initialization: Draw $\{x^i_0\}^{N_p}_{i=1}$ from the prior $p_0(x)$. Set ${\hat x}_0$ and ${\hat P}_0$  to be the mean and covariance of $p_0(x)$, respectively;
\For{$t = 1$ to $T$} 
\For {$i = 1, . . . , N_p$} 
\State  Propagate particles $\eta^i_0 = g_t(x^i_{t-1}, v_t)$;
\State  Propagate particles $\bar{\eta}^i_0 = g_t(x^i_{t-1}, 0)$;
\EndFor
\State  Calculate ensemble mean of ${\eta}^i_{0}$: ${\overline{\mu}}_{t|t-1}$;
\State  Calculate ensemble covariance matrix of ${\eta}^i_{0}$: $\overline{\Sigma}_{t|t-1}$;
\State   Apply EKF/UKF prediction:
\Statex  $\quad \quad \quad (\hat{x}_{t-1},P_{t-1|t-1}) \to (m_{t|t-1}, P_{t|t-1} )$;
\State  Set $\eta^i_1 = \eta^i_0$  and  $\theta_i = 1$, $\overline{\eta}^{i}=\overline{\eta}^{i}_{0}$;
\State  Set $\lambda$ = 0;
\For {$j = 1, . . . , N_\lambda$}
\State Set $\lambda = \lambda + \epsilon_j$;
\For {$i = 1, . . . , N_p$}
\State Calculate $A^i_j (\lambda)$ and $b^i_j (\lambda)$ from \eqref{ledh_eqA} and \eqref{ledh_eqB}
\Statex $\quad \quad \quad \quad \quad$ with the linearization being performed at $\overline{\eta}^i$;
\State  Migrate particles:
\Statex $\quad \quad \quad \quad \quad$ $\overline{\eta}^i =\overline{\eta}^i + \epsilon_j (A^i_j (\lambda) \overline{\eta}^i+ b^i_j (\lambda))$;
\State  Migrate particles:
\Statex $\quad \quad \quad \quad \quad$ $\eta^i_1 =\eta^i_1 + \epsilon_j (A^i_j (\lambda) \eta^i_1+ b^i_j (\lambda))$;
\State Calculate $\theta_i = \theta_i| det(I +\epsilon_j A^i_j (\lambda))|$;
\EndFor
\EndFor
\For {$i = 1, . . . , N_p$}
\State Compute weights $w^{i}_{t}$ using~\eqref{eq:PFGPF_weight};
\EndFor
\For {$i = 1, . . . , N_p$}
\State Normalize $w^i_t = \frac{w^i_t}{{\sum}^{N_p}_{s=1} w^s_t}$;
\EndFor
\State Compute ${\mu}_t = {\sum}^{N_{p}}_{i=1} {w}^{i}_{t}{\eta}^i_1$;
\State Compute ${\Sigma}_t = {\sum}^{N_{p}}_ {i=1} {w}^{i}_{t}({\eta}^i_1 -\mu_{t})({\eta}^i_1 -  \mu_{t})^{T}$;
\State Apply EKF/UKF update:
\Statex $\quad \quad \quad$ $(m_{t|t-1}, {P}_{t|t-1}) \to (m_{t|t}, P_{t|t})$;
\State Draw $\{x^i_t\}_{i=1}^{N_p} \sim \mathcal{N}({\mu}_t,\Sigma_{t})$;
\State State estimate: $\hat{x}_{t} = \mu_{t}$;
\EndFor
\end{algorithmic}
\end{algorithm}

\section{Simulations and results}
\label{sec:simulations}
We examine performance of the proposed PFGPF algorithm in numerical simulations of multi-target acoustic tracking and high-dimensional filtering problems. The baseline algorithms include particle flow filters such as the EDH filter ~\cite{daum2010exact} and the LEDH filter~\cite{ding2012implementation}, and particle flow particle filters such as the PFPF (EDH) and PFPF (LEDH) filters~\cite{li2017particle}.

\subsection{Multi-target acoustic tracking}
We adopt the setup for numerical simulation of multi-target acoustic tracking used in \cite{li2017particle,hlinka2011distributed}. There are $M = 4$ acoustic targets with the state evolution dynamic given by    
\begin{align}
x^m_{t} &= F x^m_{t-1} + v^m_{t} \,\,, \label{eq:acoustic_dynamic}
\end{align}
where $x^m_{t} = [{\rm x}^m_{t},{\rm y}^m_{t},\dot{{\rm x}}^m_{t},\dot{{\rm y}}^m_{t}]$ is the state of the $m$-th target, which consists of the position and velocity components of the target. The process noise $v^m_t \sim \mathcal{N}(0,V)$ is Gaussian, and the state transition matrix $F$ is given by 
\begin{align}
F = \begin{bmatrix} 
   1 & 0 & 1&0\\
   0& 1 & 0&1 \\
   0& 0& 1&0 \\
   0&0 &0 &1
  \end{bmatrix} \,\,.
\label{eq_F}
\end{align}
The targets move independently in a region of size $40 m \times 40 m$. The $s$-th sensor, located at position $r^{s}$, records the superpositional measurement 
\begin{align}
\overline{z}^s(x_t) = \sum^{M}_{m=1} \frac{\psi}{{\| {[{\rm x}^m_t,{\rm y}^m_t]}^T - r^s \|}_2 + d_0} \,\,, \label{eq_sensor_amp}  
\end{align}
where $\|\cdot\|_2$ is the Euclidean norm, $\psi = 10$ is the amplitude of the sound emitted by the targets, and $d_0 = 0.1$. There are $N_{s} = 25$ sensors located in the given region. The measurement sensed by each sensor is affected by a Gaussian Noise $\mathcal{N}(0, \sigma^2_w)$ with variance $\sigma^2_w = 0.01$, leading to highly informative measurements. The true initial states of the targets are ${[12, 6, 0.001, 0.001]}^T$, ${[32,32,-0.001,-0.005]}^T$,  ${[20, 13, -0.1, 0.01]}^T$ and ${[15, 35, 0.002, 0.002]}^T$. We have simulated 100 different trajectories with a constant velocity model and process noise covariance matrix  given by
\begin{align}
 V = \frac{1}{20}\begin{bmatrix} \label{eq_V}
   1/3 & 0 & 0.5&0\\
   0& 1/3 & 0&0.5 \\
   0.5& 0& 1&0 \\
   0&0.5 &0 &1
  \end{bmatrix}\,\,.
\end{align}  

Measurements are generated for each trajectory and each algorithm runs $5$ times on each measurement set with different initial distributions. Each initial distribution $p_0(x)$ has a mean sampled from a Gaussian with true initial states of the targets as mean and standard deviation for position and velocity components set as $10$ and $1$ respectively. The process noise covariance matrix $Q$ for the filters is set as below which assumes that there is more uncertainty during tracking.
\begin{align}
Q = \begin{bmatrix}  \label{eq_Q}
   3 & 0 & 0.1&0\\
   0& 3 & 0&0.1 \\
   0.1& 0& 0.03&0\\
   0&0.1 &0 &0.03
  \end{bmatrix} \,\,.
\end{align}
 
We run simulations with $N_{p} = 100$ and $N_{p} = 500$ particles for all the algorithms. Resampling in particle flow particle filter is performed when the effective sample size is less than $\frac{N_{p}}{2}$.

The error metric used to compare algorithmic performance is the optimal mass transfer (OMAT) metric~\cite{schuhmacher2008consistent}. The OMAT metric $d_{p}(X,\hat{X})$ is defined as 
\begin{align}
d_{p}(X,\hat{X}) =  \frac{1}{C} \left( \min_{\pi \in \Pi}  \sum^{C}_{c=1} d(x_{c},\hat{x}_{\pi(c)})^{p}  \right)^{1/p}   \,,
\end{align}
where $X = \{x_{1},x_{2},\ldots,x_{C}\}$ and $\hat{X}=\{\hat{x}_{1},\hat{x}_{2},\ldots,\hat{x}_{C}\}$ are the two sets to be compared, $p$ is a fixed scalar parameter, $\Pi$ is the set of the possible permutations of $\{1,2,...,C\}$, and $d(x,\hat{x})$ is the Euclidean distance between $x$ and $\hat{x}$. The value of $p$ is set to 1.
  
\begin{figure}[h]
\centering
\includegraphics[width=0.5\textwidth]{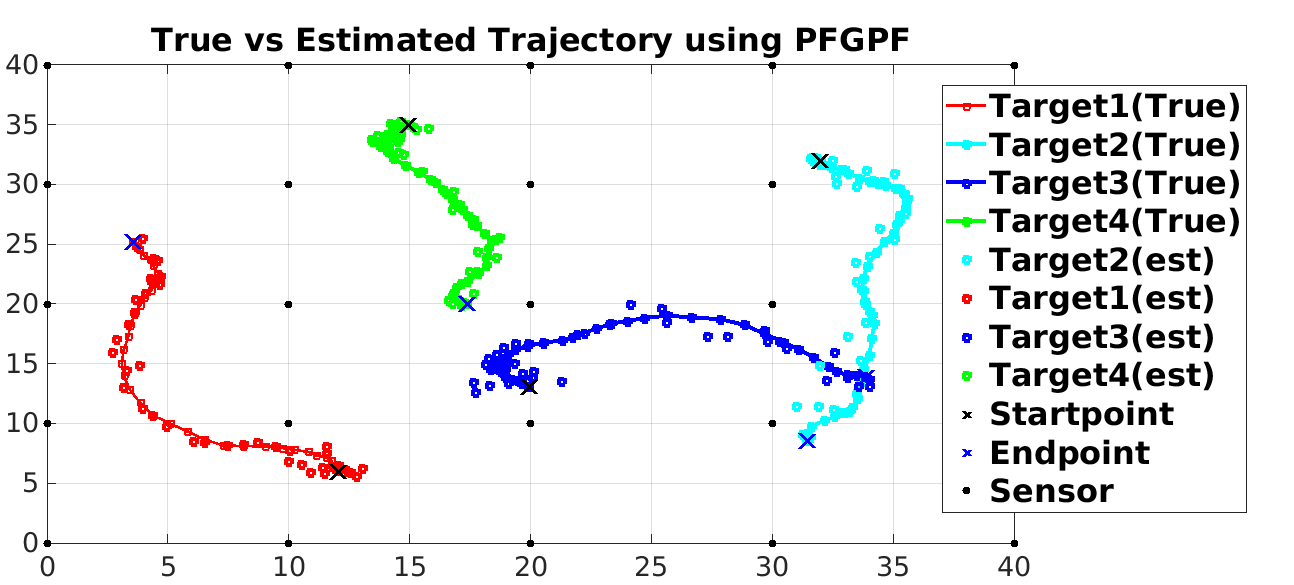}
\caption{A sample true trajectory and estimated trajectory using the PFGPF.}
\label{fig:tracks}
\end{figure}

The average OMAT error for various algorithms is given in Table \ref{tab:acoustic} for $N_p = 100$ and $N_p = 500$ particles, respectively. The proposed PFGPF has the smallest error compared to the other algorithms. 500 particles leads to significantly smaller average OMAT errors for importance sampling based methods, in particular the PFPF (LEDH) and the PFGPF, with the PFGPF leading to the smallest average tracking error. A sample plot of true trajectories and trajectories estimated using the PFGPF is shown in Figure \ref{fig:tracks}. 

\begin{table}[h]
\renewcommand{\arraystretch}{1.3}
\centering
  \caption{\label{tab:acoustic} [Acoustic model, 16 dimensions] Average OMAT error $(m)$ and average execution time $(s)$ for various algorithms.}
  \begin{tabular}{|l|l|l|l|l|}
    \hline
    \multirow{2}{*}{Algorithm} &
      \multicolumn{2}{c|}{Average OMAT (m)} &
      \multicolumn{2}{c|}{Average time (s)} \\
    \cline{2-5}
    & $N_p = 100$ & $N_p = 500$ & $N_p = 100$ & $N_p = 500$ \\
    \hline
    EDH & 2.53  & 2.61  & 0.01  & 0.01 \\
    \hline
    LEDH & 1.77 & 1.80  & 0.20  & 0.80 \\
    \hline
    PFPF (EDH) & 2.70 & 2.63 & 0.01 & 0.02 \\
    \hline
    PFPF (LEDH) & 1.25 & 0.76 & 0.38 & 1.70 \\
    \hline
    PFGPF & 1.19 & 0.73 & 0.35 & 1.60 \\
    \hline
  \end{tabular}
\end{table}

\subsection{Large spatial sensor networks: Skewed-t dynamic model and count measurements}

We evaluate the proposed algorithm in a spatial sensor network simulation setup \cite{septier2015langevin} that has been examined with the baseline algorithms~\cite{li2017particle}. The dynamic model follows Generalized Hyperbolic (GH) skewed-t distribution, a heavy-tailed distribution that is useful for modelling physical processes and financial markets with extreme behavior and asymmetric data~\cite{zhu2010generalized}. The transition kernel is given by 
\begin{align}
\begin{split}
p(x_{t}|x_{t-1}) = K_ {\frac{\nu+d}{2}}\sqrt{(\nu+Q(x_{t}))(\gamma^{T}\Sigma^{-1}\gamma)} \\
\times \frac{e^{{(x_{t}-\mu_{t})}^{T}\Sigma^{-1}\gamma}}{{\sqrt{(\nu+Q(x_{t}))(\gamma^{T}\Sigma^{-1}\gamma)}}^{-\frac{\nu+d}{2}}({1+\frac{Q(x_{t})}{\nu})}^{\frac{\nu+d}{2}}}\,\,,
\end{split}  
\label{eq_GH_density}
\end{align}
where $d$ is the number of sensors which are deployed uniformly on a two dimensional grid $\{1,2,.....,\sqrt{d}\} \times \{1,2,.....,\sqrt{d}\}$, $K_{\frac{\nu+d}{2}}$ is the modified Bessel function of the second kind of order $\frac{\nu+d}{2}$, $\mu_{t}=\alpha x_{t-1}$, $Q(x_{t})={(x_{t}-\mu_{t})}^{T}\Sigma^{-1}(x_{t}-\mu_{t})$, and the $(i,j)$-th entry of $\Sigma$ is \begin{align}
\Sigma_{i,j} = \alpha_{0}e^{\frac{{-{||R^{i}-R^{j}||}^{2}_{2} }}{\beta}}  + \alpha_{1}\delta_{i,j}\,\,, \label{eq_GH_covij}
\end{align}
where ${\|\cdot\|}_{2}$ is the Euclidean norm, $R^{i}$ is the physical position of sensor $i$, and $\delta_{i,j}$ is the Kronecker delta symbol. We set $\alpha_0 = 3, \alpha_1 = 0.01, \beta = 20$ following~\cite{septier2015langevin, li2017particle}. The shape of the distribution is defined by the parameters $\gamma$ and $\nu$. The covariance $\hat{\Sigma}$ is given by
\begin{align}
\hat{\Sigma} = \frac{\nu}{\nu-2}\Sigma + \frac{\nu^{2}}{{(2\nu-8)(\frac{\nu}{2} - 1)}^{2}}\gamma\gamma^{T} \,\,. \label{eq_GH_cov}
\end{align}
The measurements in this setup are count data which follow the Poisson distribution
\begin{align}
p(z_{t}|x_{t}) &= \prod^{d}_{c=1} \mathcal{P}_{0}(z^{c}_{t}; m_{1}e^{m_{2}x^{c}_{t}}) \,\,, \label{eq_GH_obs}
\end{align}
where $\mathcal{P}_{0}(\cdot \,; m)$ is the Poisson($m$) distribution, $m_1 =1$, and $m_2 = \frac{1}{3}$. The value of $d$ is set to 144. Each experiment is simulated for 30 time steps and we conduct the simulations 100 times. We report the results for a shorter simulation time of 10 steps as well as done in \cite{li2017particle}. $N_p = 200$ particles are used for all the algorithms. The measurement covariance $R$ in this setup depends on the state $x_{t}$, thus it is updated in each step of particle flow and before the EKF update in all algorithms. We follow the practice in \cite{septier2015langevin, li2017particle} to set the initial true state as 0 in each state dimension, as used by all the compared algorithms.  

Table~\ref{tab:septier_144} compares the mean squared error (MSE) of the algorithms used in this simulation. It has been observed in~\cite{li2017particle} that particle flow filters perform better in this simulation setup and results reported in Table~\ref{tab:septier_144} are consistent with this observation. The challenge for importance sampling-based methods in filtering in such high state dimensions is that the variance of importance samples can be high, leading to decreased performance in state estimation. 
Still, among all filters with asymptotic statistical consistency properties of the mean estimator, namely the PFPF (EDH), the PFPF (LEDH) and the PFGPF, the PFGPF exhibits the smallest average MSE.
Though the improvement in MSE of PFGPF is small compared to  PFPF (EDH) and the PFPF (LEDH) in the case of 10 time steps, the improvement is significant when 30 time steps are considered.


\begin{table}[h]
\renewcommand{\arraystretch}{1.3}
\centering
  \caption{\label{tab:septier_144}Average MSE in the large spatial sensor networks simulation. State dimension $d = 144$ and $N_p = 200$ particles.}
  \begin{tabular}{|l|l|l|l|l|}
    \hline
    \multirow{2}{*}{Algorithm} &
      \multicolumn{2}{c|}{Average MSE (m) } &
      \multirow{2}{*}{Average time (s)} \\
    \cline{2-3}
    & 10 timesteps & 30 timesteps & \\
    \hline
    EDH & 0.69 & 0.62  & 0.08   \\
    \hline
    LEDH & 0.71 & 0.64  & 10.95   \\
    \hline
    PFPF (EDH) & 0.98 & 1.09 & 0.12\\
    \hline
    PFPF (LEDH) & 0.97 & 1.08 & 22.28 \\
    \hline
    PFGPF & 0.94 & 0.87 & 21.01 \\
    \hline
  \end{tabular}
\end{table}

\section{Conclusions}
\label{sec:summary}
In this paper, we have proposed the particle flow Gaussian particle filter (PFGPF) algorithm. It embeds invertible particle flow to generate the proposal distribution within the Gaussian particle filtering framework. We explore the capacity of the (local) Gaussian approximations, introduced from both the adopted localized invertible particle flow procedure and the Gaussian particle filter model, in high-dimensional non-linear state estimation tasks. Empirical results in two challenging state estimation tasks show encouraging results compared with several particle flow filters and particle flow particle filters.
Future directions include variants based on Gaussian sum particle filters~\cite{kotecha2003GaussianSum} to further improve the filtering performance in scenarios with high-dimensional multi-modal posteriors.
%
%


\bibliographystyle{IEEEbib}
\bibliography{bib_file}

\end{document}